\documentclass[aps,prb,twocolumn,groupedaddress]{revtex4}%
\usepackage{amsmath}
\usepackage{graphics}
\usepackage{graphicx}
\usepackage{epsfig}
\usepackage{amssymb}
\usepackage{amsfonts}%
\providecommand{\U}[1]{\protect\rule{.1in}{.1in}}

\begin{document}
\title{Pressure induced superconductor quantum critical point in multi-band systems}
\author{Igor T. \surname{Padilha} and Mucio A. \surname{Continentino}}
\email{mucio@if.uff.br}
\affiliation{Instituto de F\'{\i}sica,
Universidade Federal
Fluminense, \\
Campus da Praia Vermelha, 24210-346, Niter\'oi, RJ, Brazil}

\date{\today }

\begin{abstract}
In multi-band superconductors as inter-metallic systems and heavy fermions,
external pressure can reduce the critical temperature and eventually destroy
superconductivity driving these systems to the normal state. In many cases
this transition is continuous and is associated with a superconducting quantum
critical point (SQCP). In this work we study a two-band superconductor in the
presence of hybridization $V$. This one-body mixing term is due to the overlap
of the different wave-functions. It can be tuned by external pressure and
turns out as an important control parameter to study the phase diagram and the
nature of the phase transitions. We use a BCS approximation and include both
inter and intra-band attractive interactions. For negligible inter-band
interactions, as hybridization (pressure) increases we find a SQCP separating
a superconductor from a normal state at a critical value of the hybridization
$V_{c}$. We obtain the behavior of the electronic specific heat close to the
SQCP and the shape of the critical line as $V$ approaches $V_{c}$.

\end{abstract}
\maketitle


\section{Introduction}

The study of asymmetric superconductivity, i.e., of superconductivity in
systems where different types of quasi-particles coexist at a common Fermi
surface has raised a lot of interest in the last years. This in part is due to
the relevance of this problem for many different areas in physics. It arises
in cold atomic systems with superfluid phases \cite{nature}, in color
superconductivity in the core of neutron stars \cite{6,casalbuoni,7} and in
condensed matter physics \cite{Nosso}. Furthermore it is closely related to
inhomogeneous superconductivity, as FFLO phases \cite{fflo}, since this is a
possible ground state for asymmetric systems. In condensed matter, as
inter-metallic materials, due to electrons from different orbitals at the
Fermi surface, there is a natural mismatch of their Fermi wave-vectors. This
arises because of the different effective masses of these quasi-particles or
because they occur in distinct numbers per atom.

Then, in multi-band systems, even in the absence of external magnetic fields,
one has to consider the possibility of inhomogeneous superconductivity or
other types of exotic ground states as gapless superconducting phases
\cite{liu,liu2} or phase separation \cite{caldas}.

In this paper we focus on the problem of driving a multi-band
superconductor to the normal state by applying external pressure.
When this occurs continuously, this transition is associated with a
SQCP. The theories which have been proposed for the SQCP rely in
general on the presence of disorder or magnetic impurities
\cite{sqcp}. However, there is no reason to expect that this should
play a role in clean systems driven to a normal state by external
pressure. In the model we discuss here, mixing transfers electrons
that participate in Cooper pairing to a normal band eventually
destroying superconductivity. Since there is no dissipation in the
normal band, superconductivity disappears due to a loss of coherence
in the system. Our model considers two hybridized bands in the
presence of inter and intra-band attractive interactions. These
interactions are competing and determine the nature of the zero
temperature phase transitions to the normal state as hybridization
(pressure) increases. We show that only when intra-band interactions
are dominant this transition is continuous. Otherwise it is first
order and accompanied by phase separation as usual in this case.

The problem of superconductivity in systems with overlapping bands has been
treated originally by Suhl, Matthias and Walker \cite{SMW}. The relevance of
the different interactions \cite{10} has been discussed in terms of an energy
associated with the Fermi surface mismatch, $\Delta(\delta_{k_{F}})=v_{F}
\delta_{k_{F}}$ and the critical temperature, $k_{B} T_{c}$. Only when the
former is much smaller then the latter inter-band interactions become
important. In our approach the mismatch $\delta_{k_{F}}$ depends on
hybridization and can be controlled by pressure.

\section{Model and formalism}

We consider a model with two types of quasi-particles, $a$ and $b$, with an
attractive inter-band interaction \cite{10} $\tilde{g}$, an attractive
intra-band interaction $\tilde{U}$ and a hybridization term $\tilde{V}$ that
mixes different quasi-particles states \cite{Nosso}. This one-body mixing term
$V$ is related to the overlap of the wave functions and can be tuned by
external parameters, like pressure, allowing to explore the phase diagram and
quantum phase transitions of the model. The Hamiltonian is given by
\begin{align}
&  H=\sum\limits_{k\sigma}\tilde{\epsilon}_{k}^{a}a_{k\sigma}^{\dagger
}a_{k\sigma}+\sum\limits_{k\sigma}\tilde{\epsilon}_{k}^{b}b_{k\sigma}%
^{\dagger}b_{k\sigma}\nonumber\\
&  + \tilde{g}\!\!\sum\limits_{kk^{\prime}\sigma}a_{k\prime\sigma}^{\dagger
}b_{-k\prime-\sigma}^{\dagger}b_{-k-\sigma}a_{k\sigma}+\!\tilde{U}%
\!\!\sum\limits_{kk^{\prime}\sigma}b_{k\prime\sigma}^{\dagger}b_{-k\prime
-\sigma}^{\dagger}b_{-k-\sigma}b_{k\sigma}\nonumber\\
&  +\sum\limits_{k\sigma}\tilde{V}_{k}\left(  a_{k\prime\sigma}^{\dagger
}b_{k\sigma}+b_{k\sigma}^{\dagger}a_{k\sigma}\right)  \label{eq01}%
\end{align}
where $a_{k\sigma}^{\dagger}$ and $b_{k\sigma}^{\dagger}$ are creation
operators for the light $a$ and the heavy $b$ quasi-particles, respectively.
The index $l=a,b$. The dispersion relations $\tilde{\epsilon}_{k}^{l}%
=\frac{\hbar^{2} k^{2}}{2m_{l}}-\mu_{l}$ and the ratio between effective
masses is taken as $\alpha=m_{a}/m_{b}<1$. For simplicity, we renormalize all
the energies in the problem by the chemical potential $\mu_{a}$ of the band
$a$ (the \textit{non-tilde} quantities). Furthermore we take $\hbar^{2}/(2
m_{a} \mu_{a})=1$. In this case, the dispersion relations can be written as,
$\epsilon_{k}^{a}=\tilde{\epsilon}_{k}^{a}/\mu_{a}=k^{2}-1$ and $\epsilon
_{k}^{b}=\tilde{\epsilon}_{k}^{b}/\mu_{a}=\alpha k^{2}-b$, with $b=\mu_{b}%
/\mu_{a}$.

The $V$-term is responsible for the transmutation among the quasi-particles.
In metallic systems, as transition metals\cite{11}, intermetallic compounds
and heavy fermions\cite{12}, it is due to the mixing of the wave-functions of
the quasi-particles in different orbitals through the crystalline potential.
In the quark problem, it is the weak interaction that allows the
transformation between up and down-quarks\cite{6,7,8}. For a system of cold
fermionic atoms in an optical lattice with two atomic states ($a$ and $b$),
the $V$-term is due to Raman transitions with an effective Rabi frequency
which is directly proportional to $V$ \cite{liu3}. Then, the physical origin
of the $V$-term is different for each case. The main point is that at least in
inter-metallic systems, hybridization can be easily controlled by pressure or
doping\cite{leticie} allowing to explore their phase diagram using these
quantities as external parameters. Notice that since hybridization transforms
a quasi-particle into one another, in its presence only the total number of
particles is conserved.

The order parameters that characterize the different superconducting phases of
the system described by the above Hamiltonian are, $\Delta_{ab}=-g\sum
\limits_{k\sigma}\left\langle b_{-k-\sigma}a_{k\sigma}\right\rangle $ and
$\Delta=-U\sum\limits_{k\sigma}\left\langle b_{-k-\sigma}b_{k\sigma
}\right\rangle $. These are related to inter-band and intra-band
superconductivity, respectively. The anomalous correlation functions can be
obtained from the Greens functions which also yield the spectrum of
excitations in the superconducting phases. We use the equation of motion
method to calculate standard and anomalous Greens functions \cite{16}.
Excitonic types of correlations that just renormalize the hybridization
\cite{17} are neglected. The relevant anomalous Greens functions are,
$\left\langle \left\langle a_{k\sigma};b_{-k-\sigma}\right\rangle
\right\rangle $ and $\left\langle \left\langle b_{k\sigma};b_{-k-\sigma
}\right\rangle \right\rangle $. When we write the equations of motion for
them, new Greens functions are generated \cite{16}. Some of these are of
higher order, as they contain a larger number of creation and annihilation
operators than just the two of the initials Greens functions. For these, we
apply a BCS type of decoupling \cite{16} to reduce them to the order of the
originals propagators. Finally, writing the equations of motion for the new
Greens functions, we obtain a closed system of equations that can be solved
\cite{Nosso}. The anomalous propagators from which the order parameters are
self-consistently obtained are given by \cite{Nosso}, \ \
\begin{equation}
\left\langle \left\langle a_{k\sigma};b_{-k-\sigma}\right\rangle \right\rangle
=\frac{\Delta_{ab}N(\omega)}{\omega^{4}+C_{2}\omega^{2}+C_{1}\omega+C_{0}}
\label{eq02}%
\end{equation}
with\bigskip%
\[
N(\omega)=\Delta_{ab}^{2}\!-\!V^{2}\!-\!\left(  \omega\!-\!\epsilon_{k}%
^{b}\right)  \left(  \omega\!-\!\epsilon_{k}^{a}\right)  \!+\!\frac{\Delta
}{\Delta_{ab}}\left(  \omega+\epsilon_{k}^{a}\right)
\]
and%
\begin{equation}
\left\langle \left\langle b_{k\sigma};b_{-k-\sigma}\right\rangle \right\rangle
=\frac{\Delta\left[  \left(  \omega^{2}-\epsilon_{k}^{a2}\right)
+2\frac{\Delta_{ab}}{\Delta}V\omega\right]  }{\omega^{4}+C_{2}\omega^{2}%
+C_{1}\omega+C_{0}} \label{eq03}%
\end{equation}
where%
\begin{align}
C_{2}  &  =-\left[  \epsilon_{k}^{a2}+\epsilon_{k}^{b2}+\Delta^{2}+2\left(
\Delta_{ab}^{2}+V^{2}\right)  \right] \label{eq04}\\
C_{1}  &  =4\Delta_{ab}\Delta\nonumber\\
C_{0}  &  =\left[  \epsilon_{k}^{a}\epsilon_{k}^{b}-\left(  V^{2}-\Delta
_{ab}^{2}\right)  \right]  ^{2}+\Delta^{2}\epsilon_{k}^{a2}.\nonumber
\end{align}
As mentioned before, the poles of these propagators give the excitations of
the system. Also from the discontinuity of the Greens functions on the real
axis we can obtain the anomalous correlation functions characterizing the
superconducting state. In general the appearance of exotic superconducting
phases is related to the existence of soft modes in the spectrum of
excitations \cite{liu}. In the present case, for the energy of the excitations
to vanish, it is required that $\left[  \epsilon_{k}^{a}\epsilon_{k}%
^{b}-\left(  V^{2}-\Delta_{ab}^{2}\right)  \right]  ^{2}+\Delta^{2}%
\epsilon_{k}^{a2}=0$. This can occur by tuning the hybridization parameter,
such that, $V=\Delta_{ab}$ in which case gapless excitations appear at
$k=k_{F}^{a}$ where $\epsilon_{k}^{a}=0$. Without this fine tuning there are
no gapless modes. However, in case the intra-band interaction vanishes there
is a zero energy mode for the wave-vector $k$, such that, $\epsilon_{k}%
^{a}\epsilon_{k}^{b}-\left(  V^{2}-\Delta_{ab}^{2}\right)  =0$. We will see
the effects of this behavior in the next section. If, for symmetry reasons, we
neglect the term linear in $\omega$ $\left(  C_{1}=0\right)  $, we obtain the
energy of the excitations in the form,
\begin{equation}
\omega_{1,2}(k)=\sqrt{A_{k}\pm\sqrt{B_{k}}} \label{dispersion}%
\end{equation}
with,
\begin{equation}
A_{k}=\frac{\epsilon_{k}^{a2}+\epsilon_{k}^{b2}}{2}+\Delta_{ab}^{2}%
+V^{2}+\frac{\Delta^{2}}{2} \label{eq05}%
\end{equation}
and%
\begin{align}
B_{k}  &  =\left(  \frac{\epsilon_{k}^{a2}-\epsilon_{k}^{b2}}{2}\right)
^{2}+V^{2}\left(  \epsilon_{k}^{a}+\epsilon_{k}^{b}\right)  ^{2}+\Delta
_{ab}^{2}\left(  \epsilon_{k}^{a}-\epsilon_{k}^{b}\right)  ^{2}\label{eq06}\\
&  +4V^{2}\Delta_{ab}^{2}+\frac{\Delta^{4}}{4}-\frac{\Delta^{2}}{2}\left(
\epsilon_{k}^{a2}-\epsilon_{k}^{b2}\right)  +\Delta^{2}\left(  V^{2}%
+\Delta_{ab}^{2}\right) \nonumber
\end{align}

The order parameters are determined self-consistently by a set of two coupled
equations which for finite temperatures are given by \cite{Nosso},%

\begin{equation}
\frac{1}{g\rho_{a}}\!\!=\!\!\sum\limits_{j=1}^{2}\!\!\int\limits_{-\omega_{D}%
}^{\omega_{D}}\!\!\!\frac{(-\!1)^{j}d\varepsilon}{2\sqrt{B(\varepsilon)}%
}\!\!\left[  \frac{\omega_{j}^{2}\left(  \varepsilon\right)
\!-\!\lambda^{2}(\varepsilon)}{2\omega_{j}\left( \varepsilon\right)
}\right]
\!\tanh\frac{\beta\omega_{j}\left(  \varepsilon\right)  }{2}\! \label{eq07}%
\end{equation}
with%
\begin{align}
&  \lambda^{2}\left(  \varepsilon\right)  =\left[  \frac{\varepsilon
\!+\!\left(  \alpha\varepsilon-b\right)  }{2}\right]
^{2}\!+\!\left(
\Delta_{ab}^{2}\!-\!V^{2}\right) \label{eq08}\\
&  +\!\frac{\Delta V}{4}\left[  \Delta V\!+\!4\left(  \frac{\varepsilon
+\left(  \alpha\varepsilon-b\right)  }{2}\right)  \right]  \!-\!\left[
\frac{\varepsilon\!+\!\left(  \alpha\varepsilon\!-\!b\right)  }{2}%
\!-\!\frac{\Delta V}{2}\right]  ^{2}\nonumber
\end{align}
where in $B_{k}$ and $\omega_{j}\left(  k\right)  $, we substituted
$\epsilon_{k}^{a}=\varepsilon$ and $\epsilon_{k}^{b}=\alpha+(\alpha
\varepsilon-b)$.%

\begin{equation}
\frac{1}{U\rho_{b}}\!=\!\!\sum\limits_{j=1}^{2}\!\!\!\int\limits_{-\omega_{D}%
}^{\omega_{D}}\!\!\!\frac{(\!-\!1)^{j}d\varepsilon}{2\sqrt{B(\varepsilon)}%
}\!\left[  \frac{\alpha^{2}\omega_{j}^{2}\left(  \varepsilon\right)
\!-\!\left(  \varepsilon\!+\!b\!-\!\alpha\right)  ^{2}}{2\alpha^{2}\omega
_{j}(\varepsilon)}\right]  \!\tanh\frac{\beta\omega_{j}(\varepsilon)}{2}.
\label{eq10}%
\end{equation}
In this equation, we substituted $\epsilon_{k}^{b}=\varepsilon$ and
$\epsilon_{k}^{a}=(\varepsilon+b-\alpha)/\alpha$ in $B_{k}$ and $\omega
_{j}\left(  k\right)  $. The quantities $\rho_{a}$ and $\rho_{b}$ are the
density of states at the Fermi level of the $a$ and $b$ bands and $\omega_{D}$
is an energy cut-off. The right hand sides of Eqs.\ref{eq07} and \ref{eq10}
define the gap functions $f(\Delta_{ab,}\Delta)$ and $f_{b}(\Delta_{ab,}%
\Delta)$, respectively. In the next section we discuss the behavior of the
dispersion relations, of the gap functions and obtain the free energy of the
system. From these quantities we obtain the phase diagrams for finite and zero temperatures.

\section{Results and discussions}

\subsection{Zero temperature phase diagrams for pure inter or intra-band
interactions}

For completeness we discuss briefly the behavior of the system at zero
temperature\cite{Nosso}. For purely inter-band interactions the transitions
are discontinuous and there is no SQCP in the system (see Figs. \ref{fig1} and
\ref{fig2}). From the ground state energy we can identify three characteristic
values of the hybridization. Starting from the superconducting ground state,
as hybridization increases at a value $V=V_{1}$ appears a minimum in the
ground state energy at the origin $(\Delta=0)$ that coexists with an absolute
minimum at finite $\Delta_{ab}$ associated with the superconducting state.
Further increasing $V$ there is a first order phase transition to the normal
state at $V=V_{2}$, for which the energies of the normal and superconducting
states are degenerate. For still larger $V$, the superconducting state remains
as a metastable state until $V=V_{3}$ where it stops being a minimum of the
energy. The values $V=V_{1}$, $V=V_{2}$ and $V=V_{3}$ for a fixed set of
parameters ($\omega_{D}$, $\rho_{a}$, $g$) yield zero temperature phase
diagrams shown in Ref. \cite{Nosso}. It is interesting to point out that in
the inter-band case sufficiently large values of $V$ can give rise to soft
modes in the dispersion relations which are associated with the presence of
Fermi surfaces in the superconducting state\cite{Nosso}.

For purely intra-bands interactions\cite{24}, as in the general case, the
dispersion relations of the excitations in the superconductor do not vanish
for any $k$ or $V$, since the equation $\left[  \epsilon_{k}^{a}\epsilon
_{k}^{b}-V^{2}\right]  ^{2}+\Delta^{2}\epsilon_{k}^{a2}=0$ does not present
any non-trivial solution. For $T=0$ , differently from the inter-band case, as
$V$ increases the value of $\Delta$ the minimum of the ground state energy
vanishes continuously as the system enters in the normal phase \cite{Nosso}.
There are no metastable states in this case. We also observe from the gap
equation that a minimum value of the interaction $g_{b}(V)$ is required to
sustain superconductivity. At this value of the interaction there is a
continuous second order phase transition and this mixing dependent critical
interaction $g_{b}^{c}(V_{c})$ (or interaction dependent critical
hybridization) characterizes a superconducting quantum critical point. In
practice this SQCP can be reached applying pressure to the system which is a
common procedure, for example, in the study of HF materials \cite{19}.

\subsection{General case at $T\neq0$}

In this section we consider both intra and inter-band interactions and discuss
the phase diagrams in the presence of controlled mixing and finite
temperatures. For simplicity, and to show clearly the effect of each term we
assume strong inter or intra-band terms and in each case the other interaction
is added perturbatively or neglected.

\begin{figure}[th]
\centering{\includegraphics[scale=2.2]{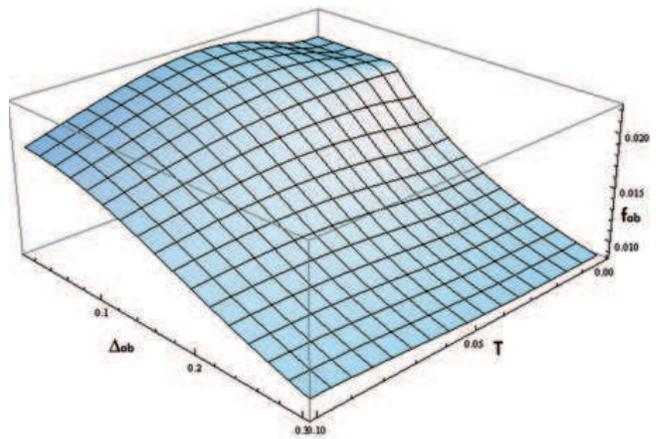}}\caption{(Color
online) 3D Graph of the asymmetrical gap function $f(\Delta_{ab},
\Delta=0)$ versus the inter-band superconducting order parameter
$\Delta_{ab}$ and temperature. See text for parameters used in the graph. }%
\label{fig1}%
\end{figure}

\begin{figure}[th]
\centering{\includegraphics[scale=0.85]{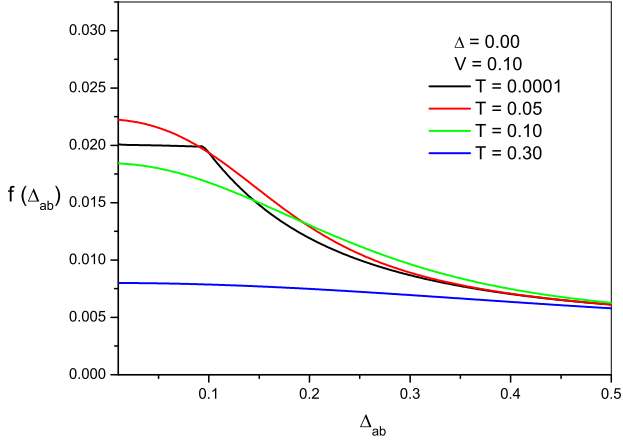}}\caption{(Color
online) Graph of the asymmetrical gap function versus the order
parameter $\Delta _{ab}$ for several values of temperature and
$\Delta=0$. These curves are intersections of the surface in Figure
\ref{fig1} with planes of
constant temperature. The parameters are the same used in Fig. \ref{fig1} (see text). }%
\label{fig2}%
\end{figure}

For dominant inter-bands interactions, such that,
$\Delta\rightarrow0$, we see in figures \ref{fig1} and \ref{fig2}
that the phase transitions that were initially discontinuous, of
first order, become continuous as temperature increases.

\begin{figure}[th]
\centering{\includegraphics[scale=0.85]{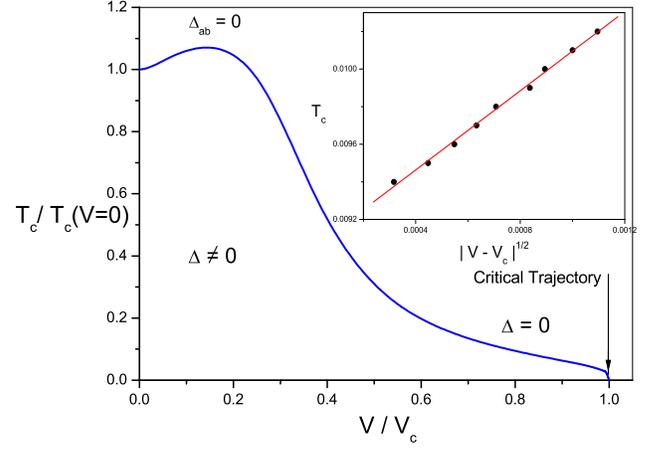}}\caption{(Color
online) Renormalized phase diagram showing  the superconducting
critical temperatures as a function of hybridization, $T_c \times
V$. The transitions along the critical line are continuous and $V_c$
is a superconducting quantum critical point. The inset shows the behavior of $T_c$
close to the SQCP. }%
\label{fig3}%
\end{figure}

Figure \ref{fig1} shows the inter-band gap function as a function
of the order parameter $\Delta_{ab}$ and temperature. Notice that
close to where the transition changes from first to second order,
there is a reentrant behavior which can also be seen in figure
\ref{fig3}. In figures \ref{fig1} and \ref{fig2} we took $\Delta=0$,
$V=0.1$, $\alpha=1/7$ and $b=0.30$. Similar values for the
parameters were used in the other figures.

We now treat the case of dominant intra-band interactions, the inter-band term
being considered perturbatively ($\Delta_{ab}\rightarrow0$) with minor
effects. In practice for inter-metallic compounds this is the case of greater
interest. Therefore, we will analyze how the usual intra-band
superconductivity changes under the influence of temperature and pressure
(hybridization). We obtain the variation of the electronic term of the
specific heat in the normal phase as hybridization changes and the system goes
through the SQCP.

Figure \ref{fig3} shows the phase diagram where the critical
temperature is plotted as a function of hybridization. The critical
line is a line of second order phase transitions. For these values
of parameters we observe that hybridization initially increases the
critical temperature before destroying superconductivity. Since the
transitions are continuous there is a SQCP at a critical value of
the hybridization $V_{c}$. For values of $V$ close to $V_{c}$, the
critical line vanishes at the SQCP as $T_{c} \propto
|V-V_{c}|^{\psi}$ with a mean-field shift exponent $\psi=1/2$ as
shown in the inset of Fig. \ref{fig3}.

In the next section we calculate the linear term of the electronic
specific heat in the normal phase as the system approaches the SQCP.

\subsection{Specific Heat}

The free energy of the system can be obtained in terms of the elementary
excitations. It is given by,%

\begin{align}
F  &  =\frac{\Delta_{ab}^{2}}{g}+\frac{\Delta^{2}}{U}+\sum_{k} f(\omega
_{1}^{0}(k),\omega_{2}^{0}(k))\\
&  -\frac{T}{2} \sum_{k} \left[  \sum_{n=1}^{2}\ln\left[  2\left(
1+\cosh\left(  \frac{\omega_{n}}{k_{B}T}\right)  \right)  \right]  \right]
\ \nonumber
\end{align}
where $\omega_{1}^{0}$ and $\omega_{2}^{0}$ are the dispersion relations for
$\Delta=\Delta_{ab}=0$, such that, the function $f(\omega_{1}^{0},\omega
_{2}^{0})$ yields the normal contribution to the free energy.

\begin{figure}[th]
\centering{\includegraphics[scale=0.85]{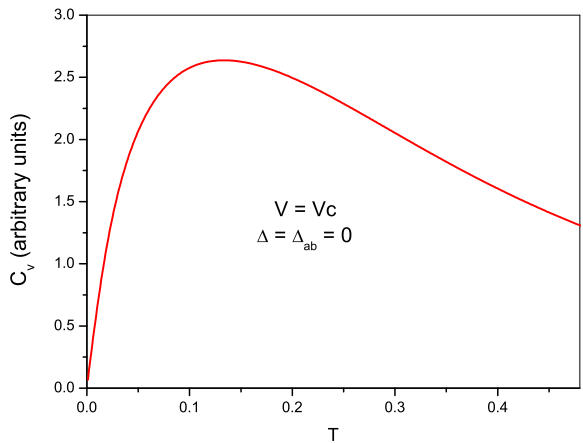}} \caption{(Color
online) Electronic specific heat as a
function of temperature along the critical trajectory ($V=V_{c}$).
The parameters used are given in the text. }%
\label{fig4}%
\end{figure}

Notice that instead of obtaining the gap functions from the Greens functions
they can be found by minimization of the free energy with respect to the order
parameters. For illustration we consider the purely inter-band case. We get,
\begin{equation}
\frac{\partial F}{\partial\Delta_{ab}}=\frac{2\Delta_{ab}}{g}+\sum_{k}%
\sum_{n=1}^{2}\left\{  -\frac{T}{2}\left[  \frac{\frac{1}{T}\frac
{\partial\omega_{n}}{\partial\Delta_{ab}}\sinh\left(  \frac{\omega_{n}}%
{k_{B}T}\right)  }{\left(  1+\cosh\left(  \frac{\omega_{n}}{k_{B}T}\right)
\right)  }\right]  \right\}  \
\end{equation}
However, $\sinh\left(  x\right)  /(1+\cosh\left(  x\right)  )=\tanh\left(
x/2\right)  $ and making $\partial F/\partial\Delta_{ab}=0$, we find
\begin{equation}
\frac{2\Delta_{ab}}{g}=\frac{1}{2}\sum_{k}\sum_{n=1}^{2}\frac{\partial
\omega_{n}}{\partial\Delta_{ab}}\tanh\left(  \frac{\omega_{n}}{2k_{B}%
T}\right)  \ .
\end{equation}
Since,
\begin{equation}
\frac{\partial\omega_{1,2}}{\partial\Delta_{ab}}=\frac{1}{2\omega_{1,2}}\left(
\frac{\partial A_{k}}{\partial\Delta_{ab}}\pm\frac{1}{2\sqrt{B_{k}}}%
\frac{\partial B_{k}}{\partial\Delta_{ab}}\right)  \
\end{equation}
and,
\begin{equation}
\frac{\partial A_{k}}{\partial\Delta_{ab}}=2\Delta_{ab}\
\end{equation}
also
\begin{equation}
\frac{\partial B_{k}}{\partial\Delta_{ab}}=2\Delta_{ab}\left[  \left(
\varepsilon_{k}^{a}-\varepsilon_{k}^{b}\right)  ^{2}+4V^{2}\right]  \
\end{equation}
we get,%
\begin{equation}
\frac{\partial\omega_{n}}{\partial\Delta_{ab}}=4\Delta_{ab}\left(
\omega _{n}^{2}-E^{2}\right)  \ ,
\end{equation}
where
\begin{equation}
E^{2}=\varepsilon_{k}^{a}\varepsilon_{k}^{b} +  \Delta_{ab}^2 - V^2
\ .
\end{equation}
Then,%
\begin{equation}
\omega_{1,2}^{2}-E^{2}=\frac{1}{2}\left[  \left(  \left(  \varepsilon_{k}%
^{a}-\varepsilon_{k}^{b}\right)  ^{2}+4V^{2}\right)  \pm\sqrt{B_{k}}\right]
\
\end{equation}
Finally, we get,%
\begin{equation}
\frac{1}{g}=\sum_{k}\sum_{n=1}^{2}\frac{\left(  -1\right)  ^{n+1}}%
{2\sqrt{B_{k}}}\left[  \left(  \frac{\omega_{n}^{2}-E^{2}}{2\omega_{n}%
}\right)  \tanh\left(  \frac{\omega_{n}}{2k_{B}T}\right)  \right]  \
\end{equation}
which is the gap equation obtained previously directly from the Greens
function. The specific heat is given by,%
\begin{equation}
C_{V}\left(  V,T\right)  =-T\left(  \frac{\partial^{2}F}{\partial T^{2}%
}\right)  \ .
\end{equation}
Using the equation for the free energy, we find,
\begin{equation}
C_{V}\left(  V,T\right)  =\frac{1}{4k_{B}}\sum_{k}\sum_{n=1}^{2}\frac
{\omega_{n}^{2}}{T^{2}}{\text{sech}}\left(  \frac{\omega_{n}}{2T}\right)  \
\end{equation}
or,%
\begin{equation}
C_{v}=\frac{\pi}{k_{B}}\sum_{n=1}^{2}\frac{1}{T^{2}}\int_{0}^{k_{F_{n}}}%
\omega_{n}^{2}\left(  k\right)  {\text{sech}}^{2}\left(  \frac{\omega
_{n}\left(  k\right)  }{2k_{B}T}\right)  k^{2}d^{3}k.
\end{equation}
This is shown in figure \ref{fig4} at the critical value of the hybridization.

\begin{figure}[th]
\centering{\includegraphics[scale=0.85]{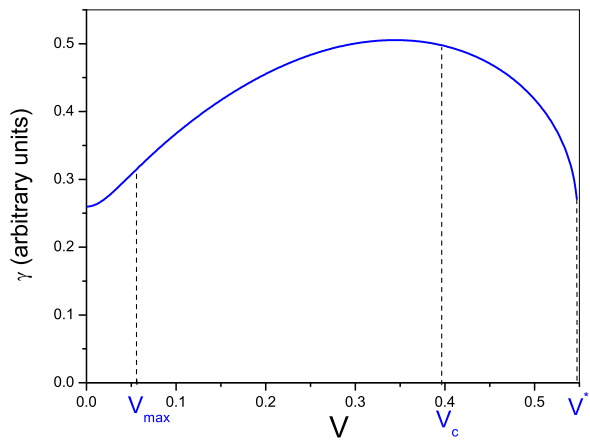}}\caption{(Color
online) Coefficient of the linear temperature dependent term of the
electronic contribution to the specific heat in the normal phase as
a function of hybridization and at $V=V_{c}$ for the same parameters
as in figure \ref{fig3}. It is also shown the value $V_{max}$ for
which $T_{c}$ is a maximum in figure \ref{fig3}. For $V=V^{*}=\tilde{V}%
^{*}/\mu_{a}=\sqrt{\mu_{b}/\mu_{a}}=\sqrt{b}$, one of the hybrid bands is
above the Fermi level and $\gamma$ has a discontinuity.}%
\label{fig5}%
\end{figure}

The linearly temperature dependent term of the electronic specific heat on the
normal phase at very low temperatures and in particular along the critical
trajectory is given by,
\begin{equation}
C_{v}=\frac{\pi^{2}}{3}k_{B}^{2}T\sum_{i=1,2}\rho_{i}\left(  \mu_{0}\right)
=\gamma T
\end{equation}
where $\rho_{i}\left(  \mu_{0}\right)  $ is the density of states of the
hybrid bands at the Fermi level. This is given by,
\begin{equation}
\rho_{i}\left(  \omega\right)  =\frac{\mathcal{V}}{\left(  2\pi\right)  ^{3}%
}\frac{4\pi k^{2}}{\left\vert \frac{\partial\omega_{i}}{\partial k}\right\vert
}%
\end{equation}
where
\begin{equation}
\frac{\partial\omega_{1,2}}{\partial k}=2k\left\{  \alpha^{+}\pm\left[
\frac{\alpha^{-}\left(  \alpha^{-}k^{2}-b^{-}\right)  }{\sqrt{\left(
\alpha^{-}k^{2}-b^{-}\right)  ^{2}+V^{2}}}\right]  \right\}
\end{equation}
with
\begin{align}
\alpha^{\pm}  &  =\frac{1\pm\alpha}{2}\\
b^{-}  &  =\frac{1-b}{2}.\nonumber
\end{align}
We wish to calculate $\rho_{i}\left(  \omega\right)  $ at the Fermi surface,
i.e., for $k=k_{F_{1,2}}$, such that, $\omega_{1}(k_{F_{1}})=0$ and
$\omega_{2}(k_{F_{2}})=0$. We finally get,%
\begin{equation}
\rho_{1,2}\left(  \omega=\mu_{0}=0\right)  =\frac{\mathcal{V}}{2\pi^{2}}%
\frac{k_{F_{1,2}}}{\left\vert \alpha^{+}\pm\left[  \frac{\alpha^{-}\left(
\alpha^{-}k_{F_{1,2}}^{2}-b^{-}\right)  }{\sqrt{\left(  \alpha^{-}k_{F_{1,2}%
}^{2}-b^{-}\right)  ^{2}+V^{2}}}\right]  \right\vert }%
\end{equation}
The values of $k_{F_{1,2}}$ can be easily obtained and we get the results for
the coefficient $\gamma(V)$ of the linear term of the specific heat as a
function of hybridization shown in Figure \ref{fig5}. We have used the same
set of parameters which yield the phase diagram shown in Figure \ref{fig3}.
From these figures we notice that the maximum $T_{c}$ occurs in a region of
the phase diagram where $\gamma$ is increasing with $V$. At the critical value
of the hybridization the coefficient of the linear term is passing through a
broad maximum. The values of $\gamma$ can be measured all along in the normal
phase, above $T_{c}$, as hybridization is increasing with external pressure
and passes through the superconducting quantum critical point at $V_{c}$. A
behavior of $\gamma$ as a function of pressure as that shown in Figure
\ref{fig5} would be a strong indication that the present mechanism is
responsible for destroying superconductivity. Since $\gamma$ is proportional
to the total density of states at the Fermi level Figure \ref{fig5} is helpful
to understand the initial increase of $T_{c}$ with hybridization as shown in
Figure \ref{fig3}. In the BCS approximation used here this is related to the
density of states at the Fermi level and this, as shown by the behavior of
$\gamma$, increases initially as $V$ increases.

The results of this paper are obtained using a mean field
approximation which does not include fluctuations. The mean-field
character of the theory is reflected, for example, in the shape of
the critical line that vanishes as $T_{c}\propto\sqrt{|V-V_{c}|}$
close to the SQCP as shown in Figure \ref{fig3}. The specific heat
calculated above was obtained in the normal phase and is due solely
to the contributions of unpaired quasi-particles in the hybridized
bands. There are no effects of fluctuations since they are not taken
into account.

Due to the nature of the approximations we used, we can expect that the
superconducting temperatures reflects the variations in the density of states
and consequently in $\gamma$ as hybridization is changed. As pointed out
before the increase in $T_{c}$ for small $V$ is entirely consistent with the
behavior of $\gamma$ shown in Figure \ref{fig5}. What is remarkable however is
that superconductivity is destroyed while still both bands contribute to the
density of states at the Fermi level. It can be easily shown that $k_{F_{2}%
}^{2}-k_{F_{1}}^{2}=2\sqrt{\left(  \frac{\alpha-b}{2\alpha}\right)  ^{2}%
+\frac{V^{2}}{\alpha}}$, such that, hybridization increases the mismatch
$\delta k_{F}=|k_{F_{1}}-k_{F_{2}}|$ between the Fermi wave-vectors of the
hybrid bands. As It is well known\cite{caldas} this can give rise to
superconducting instabilities which however always occur discontinuously
through first order transitions. In the present case the $T=0$
superconductor-normal transition is continuous being associated with a SQCP.
Since there is no dissipation in the electronic bands but a lack of coherence
in one of them we may attribute to this the destruction of the superconducting phase.

It is clear that in inter-metallic systems hybridization occurs even
at zero pressure. The point we wish to emphasize is that this
depends on applied pressure and for this reason it can be used as a
control parameter. A final remark is that as a matter of fact the
control parameter is $V/\mu_{a}$ which is the ratio of hybridization
over the bandwidth of the light quasi-particles. The latter also
depends on pressure and although we have naturally assumed that the
ratio \textit{increases} with pressure, this is not immediate and
may depend on the particular system.

We have investigated a mechanism to drive a multi-band
superconductor to a superconducting quantum critical point through
the application of pressure or doping\cite{leticie}. It does not
rely on the presence of magnetic impurities or disorder, but on the
sensitivity of hybridization to these external parameters. Evidence
that the mechanism we are proposing is in action can be obtained
from measuring the coefficient of the linear term of the electronic
specific heat just above the superconducting transition as a
function of pressure. This shows features which can be correlated
with the behavior of $T_{c}$. Superconductivity is destroyed when
all hybrid bands still contribute to the density of states and the
zero temperature transition is from the superconductor to a metallic
state.

{\acknowledgements We wish to thank the Brazilian agencies, FAPEAM,
FAPERJ and CNPq for financial support to this work and Heron Caldas
for useful discussions.}

\end{document}